\documentclass[12pt]{article}
\usepackage[all]{xy}
\usepackage{amssymb} 
\usepackage{amsmath} 
\usepackage{pdfsync}                         
\usepackage{color} 
\font\frak=eufm10 scaled\magstep1
\def\goth #1{\hbox{{\frak #1}}}
\def\<#1>{\langle#1\rangle}

\def\pd#1#2{\frac{\partial#1}{\partial#2}}

\def\X{{\goth X}}
\def\V#1{\overrightarrow{\kern-2pt#1\kern0pt}}  

\def\G{{\Gamma}}

\def\W{\Omega}

\def\be{\begin{equation}}
\def\ee{\end{equation}}
\def\ba{\begin{eqnarray}}
\def\ea{\end{eqnarray}}
\def\ben{\begin{equation*}}
\def\een{\end{equation*}}
\def\ban{\begin{eqnarray*}}
\def\ean{\end{eqnarray*}}

\newtheorem{theorem}{Theorem}

\begin{document}
\renewcommand\theequation{\arabic{section}.\arabic{equation}}
\catcode`@=11 \@addtoreset{equation}{section}

\title{Jacobi multipliers and Hojman symmetry}

\author{Jos{\'e}\ F.\ Cari{\~n}ena and Manuel F. Ra\~nada\\
\small Departamento de F\'{\i}sica Te\'orica and IUMA,
Universidad de Zaragoza.\\ 
\small  50009, Zaragoza, Spain\\
\small  e-mail: jfc@unizar.es, mfran@unizar.es\\
 }
 \medskip
 
\maketitle

\begin{abstract}
\noindent 
The geometric intrinsic approach to Hojman symmetry is  developed and use is made of the theory of the Jacobi last multipliers to find the corresponding conserved quantity for non divergence-free vector fields. The particular cases of autonomous  Lagrangian and Hamiltonian systems are studied as well as the generalization of these results to normalizer vector fields 
of the dynamics.  The nonautonomous cases, where normalizer vector fields play a relevant role,  are also developed.
\end{abstract}

\bigskip

 \medskip
 
\medskip


 \medskip

\pagebreak

\section{Introduction}

The first-integrals of a vector field in a $n$-dimensional manifold $M$, $X={\displaystyle  \sum_{i=1}^nX^i(x) \partial/\partial x^i}$, play a very relevant role in its integrability, and therefore in that of  the corresponding autonomous system of first-order differential equations:
\begin{equation}
\dot x^i=X^i(x),\quad i=1,\ldots,n.\label{autsyst}
\end{equation}
Moreover, the knowledge of other  tensor fields that are invariant under the given vector field furnishes tools for reduction to simpler cases, and the best known example 
  is the case of an  invariant symplectic form $\omega$,  for which  one of the more celebrated related results is that of the  
  Noether theorem,  that in this case amounts to say that if a Hamiltonian vector field $X_F$
  is a symmetry of the Hamiltonian function $H$, then the function $F$ is a constant of the motion for the dynamical Hamiltonian vector field $X_H$. The very simple 
  proof is based on the skew-symmetry of the  Poisson bracket defined by the symplectic structure, because $X_HF=\{F,H\}=-X_FH$.  The particular case of 
  regular Lagrangians  can be considered as a particular case of a Hamiltonian system where, in this case,  the symplectic form $\omega_L$  and the Hamiltonian function $H$, 
   the energy function,  are defined from the Lagrangian $L$. The symmetries of both the symplectic form and  the energy  function correspond to the invariance of the Lagrangian,
   may be up to  addition of a gauge term. Just to mention another instance, the existence 
  of an invariant volume form for a vector field $X$ in an oriented manifold, which is equivalent to know a Jacobi multiplier for $X$, was used by  Jacobi to prove that if $(n-2)$ 
  first integrals are known,  the vector field is integrable by quadratures.
  
  Even if the usual way for finding first-integrals  is the method suggested by  Noether theorem of introducing symmetries, there is an alternative method for finding
   non-Noether constants of motion that  is not based on symmetries but on the existence of alternative compatible  geometric structures for the 
    description of the vector field, what leads to the existence of a recursion operator. 
  The case of the Lagrangian formalism  for a   system with a one-dimensional configuration space was given in \cite{CS66}, while the multidimensional case was given in
  \cite {HH81} and the geometric approach in intrinsic terms was given in \cite{CI83}.
  
  The third approach we want to study here is the one started  in a coordinate dependent way by Hojman \cite{SH92a} for the case of a divergence-free non-autonomous system of 
  second-order ordinary differential equations  and later on extended by  Gonz\'alez-Gasc\'on in \cite{FGGJPA27} for a non divergence-free case, and that it  is becoming more and more 
  important during   the last years for its applications in $f(R)$-gravity and FRW cosmology \cite{DGR20,DGR21,WZLZ15,WZLZ16,CR13,PC15,PLC16}. 
  The geometric approach to autonomous systems provides a method to incorporate holonomic constraints by replacing the systems by vector fields in a differentiable manifold and  
  allows to develop coordinate free formulations of the problem. In this way the geometric theory of reduction and symmetry may be very helpful. Therefore, our aim in this paper is to 
  present a   geometric approach generalising the previously known cases and unifying them in a common simple geometric formulation, by showing that  they appear  as particular cases 
  of a more general theory.
   
  The paper is organized as follows:  Section 2 summarises basic results of the differential geometry concerning the concept of divergence of a vector field and a basic relation
   between the divergences of two vector fields, and we also recall the fundamental concept of Jacobi multiplier introduced by him to integrate by quadratures a vector field when
   $(n-2)$ functionally independent  first-integrals and a Jacobi multiplier are known. Using these concepts we can introduce some mathematical results which can be used  to obtain 
   a  first-integral of the vector field $X$  in case that an infinitesimal  symmetry $Y$ of $X$ or a normalizer vector field are known. After this geometric presentation we particularise
   in Section 3  to the cases of autonomous Lagrangian and Hamiltonian systems,  while  nonautonomous systems of first-order differential equations are analysed in Section 4, 
   and in particular the Hamilton systems. Section 5 is devoted to nonautonomous systems of second-order differential equations with a special emphasis on those 
   admitting a  regular Lagrangian formulation. Many results on constants of motion scattered by the physics literature arise here as particular cases
    of this more general  unifying presentation.

\section{First integrals derived from infinitesimal symmetries  of vector fields}

Different relations among vector fields and their divergences can be used to establish first integrals and integral invariants for vector fields in a manifold $M$.
Given  a vector field $X\in\mathfrak{X}(M)$ in an oriented manifold $(M,\Omega)$, we denote ${\rm div}(X)$ the function defined by
 $\mathcal{L}_X\Omega ={\rm div}(X)\,\Omega$. In the particular case of a choice of local coordinates $(x^1,\ldots, x^n)$ such that  $\Omega=dx^1\wedge \cdots\wedge dx^n$, if  the local expression of $X$ is $X={\displaystyle \sum_{i=1}^nX^i\pd{}{x^i}}$, then the expression of ${\rm div} (X)$ is the usual one: 
 $${\rm div} (X)=\sum_{i=1}^n  \pd{X^i}{x^i}.
 $$
 
 Note that for each function $f\in C^\infty(M)$, from  $\mathcal{L}_{f\,X}\Omega=
 \mathcal{L}_{X}(f\, \Omega)=(X(f)+f\, {\rm div}(X))\Omega$  we have that 
 ${\rm div}(fX)= X(f)+f\, {\rm div}(X)$. 
 
 There is a geometric relation playing a fundamental r\^ole:  If $X, Y$ is an arbitrary pair of vector fields in an oriented manifold $(M,\Omega)$, then
\begin{equation}
\mathcal{L}_X({\rm div\,} (Y))-
\mathcal{L}_Y({\rm div\,} (X))={\rm div\,} ([X,Y]), \label{reldosdiv}
\end{equation}
because as  the Lie derivatives of a $p$-form $\alpha$  satisfy the relation 
 \begin{equation}
\mathcal{L}_X(\mathcal{L}_Y\alpha)-\mathcal{L}_Y(\mathcal{L}_X\alpha)=\mathcal{L}_{[X,Y]}\alpha, \label{LXdg0}
\end{equation}
in the particular case  $\alpha=\Omega$, we have
\begin{equation}
\mathcal{L}_X(\mathcal{L}_Y\Omega)-\mathcal{L}_Y(\mathcal{L}_X\Omega)=\mathcal{L}_{[X,Y]}\Omega, \label{LXdgW}
\end{equation}
and from the difference of $$\mathcal{L}_X(\mathcal{L}_Y\Omega)=\mathcal{L}_X({\rm div}(Y)\,\Omega)=\mathcal{L}_X({\rm div}(Y))\,\Omega+{\rm div}(Y)\,{\rm div}(X)\,\Omega$$ and the corresponding 
relation  $$\mathcal{L}_Y(\mathcal{L}_X\Omega)=\mathcal{L}_Y({\rm div}(X)\,\Omega)=\mathcal{L}_Y({\rm div}(X))\,\Omega+{\rm div}(X)\,{\rm div}(Y)\,\Omega$$ we obtain (\ref{reldosdiv}).

 The following theorem makes use of this geometric property and generalises some previous results which had been  obtained in  \cite{SH92a,FGGLNC19} by making use of a coordinate dependent approach:

\begin{theorem}\label{divycom}
Let $X$ be a divergence-free vector field and $Y$ an infinitesimal symmetry of $X$. 
Then ${\rm div\,}(Y)$ is a constant of the motion for $X$.
\end{theorem}                                                          
{\sl Proof.-}    Let us apply the relation  (\ref{reldosdiv}) to a divergence-free vector field $X$.
Then if the  vector field   $Y\in\mathfrak{X}(M)$ is a symmetry of $X$,   
it satisfies $[Y,X]=0$ and therefore  we obtain that  $\mathcal{L}_X({\rm div}(Y))=0$.

\hfill$\Box$

Next, in the following two theorems, we present two generalizations of this result. 
First, we delete  the condition for $X$ to be a divergence-free vector field and  second,  the condition  of $Y$ being an exact symmetry is substituted for the weaker condition of symmetry of the one-dimensional distribution generated by  $X$. So, if the condition for $X$ to be  a divergence-free vector field   is eliminated we can use the Jacobi multiplier  theory and then we   have a more general 
result obtained in \cite{FGGJPA27,Lutzky95}.  Recall first that a positive function $R\in C^\infty(M)$ is a Jacobi multiplier for $X\in\mathfrak{X}(M)$ in the oriented 
manifold $(M,\Omega)$ if the vector field
$R\, X$ is such that $\mathcal{L}_{R\, X}\Omega=0$, or equivalently, as $\mathcal{L}_{R\,X}(\Omega)=d(i(R\,X)\Omega=d(i(X)(R\,\Omega))=\mathcal{L}_X(R\,\Omega)$,  if $\mathcal{L}_X(R\,\Omega)= 0$. Note that
$$\mathcal{L}_X(R\,\Omega)=\mathcal{L}_X(R)\,\Omega+R\,\mathcal{L}_X\Omega=(\mathcal{L}_X(R)+R\,{\rm div}(X)) \Omega,$$
and therefore the condition for $R$ to be a Jacobi multiplier  (see  \cite{CS21} and references therein for more details) becomes
\begin{equation}
\mathcal{L}_X(R)+R\,{\rm div}(X)=0\Longleftrightarrow {\rm div}(X)+\mathcal{L}_X(\log R)=0.\label{JMcond}
\end{equation}

Remark that from the above mentioned property $\mathcal{L}_{X}(R\,\Omega)=\mathcal{L}_{R\, X}\Omega$,
we see that $R$ is a Jacobi multiplier for $X$ if and only if the vector
field $X$ is  divergence-free with respect to $R\,\Omega$,  i.e., we see   the equivalence of searching for Jacobi multipliers on one side and   the more geometric concept of invariant volume-forms on the other: a volume form $\Omega'=R\,\Omega$
is $X$-invariant, i.e.  $\mathcal{L}_{X}\Omega'=0$,  if and only if $R$ is a  Jacobi multiplier for the vector field $X$ in the oriented manifold $(M,\Omega)$. Moreover, this property 
also  shows that $R$ is a Jacobi multiplier for $X$ in the oriented manifold $(M,\Omega)$ iff and only if $f\,R$   is a Jacobi multiplier for $X$  in the oriented manifold 
$(M,f^{-1}\,\Omega)$, for each positive function $f$.

\begin{theorem}\label{hojman}
 If the function  $R$ is a Jacobi multiplier  for the vector field $X$ and the vector field $Y$ is an infinitesimal symmetry of $X$, i.e. $[X,Y]=0$, then the function 
\begin{equation}
I={\rm div\,}(Y)+Y(\log R)\label{fiI}
\end{equation}
is a constant of the motion for $X$.
\end{theorem}
{\sl Proof.-}  If  $[X,Y]=0$,  then according to the relation  (\ref{reldosdiv}) for such a  pair $X,Y,$ of vector fields we have that $\mathcal{L}_X{\rm div\,} (Y)=
\mathcal{L}_Y{\rm div\,} (X)$. If  $R$ is a Jacobi multiplier for the vector field $X$,  then  (\ref{JMcond}) shows that  $\textrm{div}(X)=-\mathcal{L}_X(\log R)$, 
and consequently in this case the equation  (\ref{reldosdiv}) can be rewritten as  
$$\mathcal{L} _X{\rm div\,} (Y)=-\mathcal{L}_Y(\mathcal{L}_X(\log R))=-\mathcal{L} _X(\mathcal{L}_Y(\log R)).
$$
Therefore, $\mathcal{L}_XI=0$. 

\hfill$\Box$

Another important case is when the vector field $Y$ is not a symmetry of the vector field $X$ but only of the one-dimensional distribution spanned by $X$, i.e. there exists a
 function $h\in C^\infty(M) $ such that $[Y,X]=h\, X$. In this case the results of the previous theorems can be generalised to:
 
\begin{theorem}\label{divycomR}
i) If  the vector field $X$  is divergence-free and the vector field $Y$ is an infinitesimal symmetry of the one-dimensional distribution generated by  $X$, i.e. $[Y,X]=
h\, X$, then  the function $\textrm{div\,}(Y)+h $
is a constant of the motion for $X$.

 ii)  If the function  $R$ is a Jacobi multiplier  for the vector field $X$ and the vector field $Y$ is an infinitesimal symmetry  of the one-dimensional distribution 
generated by $X$, i.e.  i.e. there exists a
 function $h\in C^\infty(M) $ such that $[Y,X]=h\, X$, then the function 
\begin{equation}
I={\rm div\,}(Y)+Y(\log R)+h\label{fiI2}
\end{equation}
is a constant of the motion for $X$.
\end{theorem}
{\sl Proof.-}  {\it i)} If ${\rm div\,}(X)=0$, then $\mathcal{L}_X\Omega=0$, and hence, using (\ref{reldosdiv}), 
$$\mathcal{L}_X(\mathcal{L}_Y\Omega) =\mathcal{L}_Y(\mathcal{L}_X\Omega)+\mathcal{L}_{-h\,X}\Omega=-\mathcal{L}_X(h\,\Omega)=-X(h)\,\Omega,
$$
and therefore,  from 
$$\mathcal{L}_X(\mathcal{L}_Y\Omega) =\mathcal{L}_X({\rm div\,}(Y)\,\Omega)=\mathcal{L}_X({\rm div\,}(Y))\Omega
$$
we obtain:
$$
\mathcal{L}_X({\rm div\,}(Y)+h)=0.
$$

{\it ii)} Using the relation (\ref{reldosdiv}) for the vector fields $X$ and $Y$,
and the Jacobi multiplier condition ${\rm div\,} (X)+X(\log R)=0$, together with $[Y,X]=h\, X$, we find that
$$\mathcal{L}_X{\rm div\,} (Y)-\mathcal{L}_Y(-\mathcal{L}_X(\log R))=-{\rm div\,} (h\,X)=-X(h)-h\,{\rm div\,} (X),
$$
and as the commutator $[\mathcal{L}_X,\mathcal{L}_Y] $ is $\mathcal{L}_{-h\,Y}$, the preceding relation can be rewritten as
$$\mathcal{L}_X({\rm div\,} (Y)+Y(\log R))+\mathcal{L}_{h\,X}(\log R)=-X(h)+h\, X(\log R)
$$
and simplifying terms,
$$\mathcal{L}_X({\rm div\,} (Y)+Y(\log R)+h)=0.
$$
\hfill$\Box$

Given a vector field $X\in \mathfrak{X}(M)$ the vector fields  $Y\in \mathfrak{X}(M)$   that are symmetries of the distribution generated by $X$ are sometimes called normalizers of $X$.

 With the particular choice of local coordinates $(x^1,\ldots, x^n)$ such that  $\Omega=dx^1\wedge \cdots\wedge dx^n$, if  the local expressions of the vector fields
  $X$ and $Y$ are  $$X=  \sum_{i=1}^nX^i\pd{}{x^i},\qquad Y=  \sum_{i=1}^nY^i\pd{}{x^i},$$ 
 the coordinate expression of constant of motion (\ref{fiI2}) is
 $$I= \sum_{i=1}^n \pd{Y^i}{x^i}+ \sum_{i=1}^nY^i\pd{(\log R)}{x^i}+h\,, 
 $$
 and the condition $[Y,X]=h\, X$ means that
 $$
 Y(X^i)-X(Y^i)=h\, X^i.
 $$

\section{Hojman symmetry in the  Lagrangian  and Hamiltonian formalism}

We can extend the preceding results to  systems of second-order differential equations in normal form,
\begin{equation}
\ddot x^i=F^i(x,\dot x),\quad i=1,\ldots,n,\label{autsyst2nd}
\end{equation}
as done in \cite{FGGJPA27,Lutzky95}, and the particular case in which this system is equivalent to 
 the set of Euler-Lagrange equations of a regular Lagrangian is of a special interest. 
 
 Remark that such a system of $n$ second-order differential equation can be associated with a system of $2n$ first-order differential equations 
 \begin{equation}
 \left\{\begin{array}{rcl} \dot x^i&=& v^i\\ \dot v^i&=& F^i(x,v)\end{array}\right. ,\quad i=1,\ldots,n, \label{autsyst2nd2}
 \end{equation}
 that as a particular case of (\ref{autsyst}) has associated the vector field 
  \begin{equation}
  \Gamma=\sum_{i=1}^n \left(v^i \pd{}{x^i}+F^i(x,v) \pd{}{v^i}\right). \label{SODEvf}
 \end{equation}
 
When the configuration space $Q$ of the system is a $n$-dimensional Euclidean space
  the  Euclidean coordinates $(x^1,\ldots,x^n)$   induce global coordinates in its tangent bundle, denoted  $(x^1,\ldots,x^n, v^1,\ldots,v^n)$
and there is a natural volume element (tangent and cotangent bundles can be  identified by the Euclidean metric) given by 
$\Omega=dx^1\wedge\cdots\wedge dx^{n}\wedge dv^1 \wedge\cdots\wedge dv^{n}$. If the dynamical vector field (\ref{SODEvf}) 
 is determined by the Lagrangian $L$, i.e. $i(\Gamma)\omega_L=dE_L$ (see e.g \cite{CIMM}),  we 
 know (see e.g. \cite{CS21} and references therein) that 
 a particular Jacobi multiplier for given by the determinant of the Hessian matrix in the velocities, with elements 
$W_{ij}=\partial^2 L/{\partial v^i\partial v^j}$. Therefore, the constant of motion  obtained in \cite{Lutzky95} is just  a particular case of the expression  (\ref{fiI}) for $R$ equal to the 
determinant of the Hessian matrix $W$. In the more general case of $Q$ a $n$-dimensional manifold, a local chart of coordinates $(x^1,\ldots,x^n)$  for $Q$  induces a local chart of coordinates in its tangent bundle, denoted  $(x^1,\ldots,x^n, v^1,\ldots,v^n)$, and an 
associated  volume element  in such a chart given by 
$\Omega=dx^1\wedge\cdots\wedge dx^{n}\wedge dv^1 \wedge\cdots\wedge dv^{n}$. If the dynamical vector field $\Gamma$  given in such coordinates by  (\ref{SODEvf}) 
 is determined by the Lagrangian $L$, i.e. $i(\Gamma)\omega_L=dE_L$, which implies  $\mathcal{L}_\Gamma\omega_L=0 $ (see e.g \cite{CIMM}),  we 
 know (see e.g. \cite{CS21} and references therein) that 
 a particular Jacobi multiplier for $\Gamma$ with respect to the volume form $\Omega$ is  given by the determinant of the Hessian matrix $W$ in the velocities, with elements 
$W_{ij}=\partial^2 L/{\partial v^i\partial v^j}$.  Actually, from $\mathcal{L}_\Gamma\omega_L=0 $, we see that  $(\omega_L)^{\wedge n}$ is an invariant volume under $\Gamma$, and the proportionality factor of $(\omega_L)^{\wedge n}$ and $\Omega$ is $\det W$.  Therefore, the constant of motion  obtained in \cite{Lutzky95} is just  a particular case of the expression  (\ref{fiI}) for $R$ equal to the 
determinant of the Hessian matrix $W$.

 Let us now consider the coordinate expression of the vector  field $Y$  defined in $TQ$ with local coordinates $(x^i,v^i)$ 
    \begin{equation}\label{cevfY}
 Y=\sum_{i=1}^n\left(Y^i(x,v)\pd{}{x^i} +{\bar Y^i}(x,v)\pd{}{v^i}\right),
    \end{equation}
 where the fuctions $\bar Y^i$ are given by $\bar Y^i=\Gamma(Y^i)$, because of the symmetry condition $[\Gamma,Y]=0$. 
 Then the coordinate expressions of the constants of motion $I$ given in Theorem \ref{divycom}
 and in (\ref{fiI}) of Theorem  \ref{hojman} for the vector field $\Gamma$  are given, respectively, by 
 \begin{equation}
 I=\sum_{i=1}^n\left(\pd{Y^i}{x^i} +\pd{\bar Y^i}{v^i}\right),\label{IdivY}
\end{equation}
 and 
   \begin{equation}
 I=\sum_{i=1}^n\left(\pd{Y^i}{x^i} +\pd{\bar Y^i}{v^i}\right)+\frac 1R \sum_{i=1}^n\left(Y^i\pd{R}{x^i} +{\bar Y^i}\pd{R}{v^i}\right),\label{IdivYR}
   \end{equation}
 These expressions coincide with those of \cite{DGR20, DGR21}. 
 
 Similarly,  let us now consider the time  evolution in phase space given in local coordinates $(q^1,\cdots,q^n, p_1,\cdots,p_n)$ of  the cotangent bundle $\pi:T^*Q\to Q$ induced from a chart $(q^1,\ldots,q^n)$ in the base $Q$ by
\begin{equation}
    \left\{\begin{array}{rcl} \dot q^i&=&Q^i(q,p) \\ \dot p_i&=&P_i(q,p)\end{array}\right.,\label{asinps}
    \end{equation}
 whose solutions  correspond to the integral curves of the vector field 
 $$  X=\sum_{i=1}^n\left(Q^i(q,p) \,\pd{}{q^i} + P_i(q,p)\,\pd{}{p_i} \right)\,. 
 $$ 
 and let $\Omega$ be the volume form induced from exterior product of its  natural symplectic structure, 
 $\Omega=dq^1\wedge\cdots dq^n \wedge dp_1\wedge\dots dp_n$. Recall  that the cotangent bundle is the local prototype of an exact symplectic manifold. There is a canonical
  1-form $\theta\in \bigwedge^1(T^*Q)$ defined by $\theta_\alpha= \pi ^*_\alpha\alpha$, with local expression  $\theta={\displaystyle\sum_{i=1}^n}p_i\, dq^i$. Then $\omega=-d\theta$ is a symplectic form on $T^*Q$ and the local coordinates are Darboux coordinates.

 Then if $R$  is a Jacobi multiplier for $X$,  satisfying therefore the equation  (\ref{JMcond}), 
 $$\sum_{i=1}^n\left(\pd{Q^i}{q^i}+\pd{P_i}{p_i}\right)+X(\log R)=0 ,
 $$ 
 and the vector field $Y$, given in these   local coordinates by
 $$ Y={\displaystyle \sum_{i=1}^n}\left(\xi^i(q,p) \pd{}{q^i}  + \eta_i(q,p)\, \pd{}{p_i}\right) ,
 $$ 
 is a symmetry of the vector field $X$,  the constant of motion (\ref{fiI}) for $X$ is in this case
 $$I= \sum_{i=1}^n\left(\pd{\xi^i}{q^i}+\pd{\eta_i}{p_i}+\xi^i\pd{(\log R)}{q^i}+\eta_i\pd{(\log R)}{p_i}\right),
 $$
 as found in \cite{mei}. Of course, if $(M,\omega)$ is a symplectic manifold and $X$ is a Hamiltonian vector field and we consider the volume form in $M$ obtained by exterior products of the symplectic form, as $X$ is divergence-free, the multiplier $R$ is just a constant. 
 
 Finally, if the vector field $Y$ 	is a normalizer for $X$, i.e., there exists a fuction $h\in C^\infty(M)$ such that $[Y,X]=h\, X$, then the constant of motion corresponding   to that of Theorem \ref{divycomR} is 
  $$I= \sum_{i=1}^n\left(\pd{\xi^i}{q^i}+\pd{\eta_i}{p_i}+\xi^i\pd{(\log R)}{q^i}+\eta_i\pd{(\log R)}{p_i}\right)+h.
 $$
  \section{Hojman symmetry for nonautononous systems of  first-order differential equations}
   
 In the geometric approach to  non-autonomous systems of first-order differential equations,
\begin{equation}
 \dot x^i=X^i(t,x),\quad i=1,\ldots,n, \label{nautsyst}
 \end{equation}
  time is a new coordinate and then we can change the parameter 
 of the solution curves. This amounts to consider the curves not as parametrized curves but simply as one-dimensional submanifolds and to consider as equivalent two curves that are related by a reparametrization. Therefore
 we associate with such a non-autonomous system a one-dimensional, and therefore integrable, distribution,  generated either by a vector field $X$  in $\mathbb{R}\times  M$, or by any other proportional one, $f\, X$, with $f$ a nonvanishing function, 
 in such a way that its one-dimensional integral  manifolds  are the equivalence class of integral curves of a representative of the vector field. We can choose, for instance 
 \begin{equation}\label{repvf}
 X=\pd{}t+ \sum_{i=1}^nX^i(t,x)\pd{}{x^i},
  \end{equation}
  if for a vector field of the distribution we have $\<dt,X>\ne 0$. 

The concept of infinitesimal symmetry must be changed and we say that the vector field $Y\in\mathfrak{X}(\mathbb{R}\times  M)$ is a symmetry when the image under
 its local flow of an integral curve of $X$ 
is, up to a reparametrization, an integral curve of $X$. This condition means that there exists a function  $\lambda\in C^\infty (\mathbb{R}\times  M)$ such that 
$[Y,X]= \lambda\, X$, or in other terms, that the  one-dimensional distribution generated by $X$ is invariant under the flow of the vector field $Y$. Note that if the coordinate
 expression of the vector field $Y$ is 
$$Y=Y^0(t,x)\pd{}t+ \sum_{i=1}^nY^i(t,x)\pd{}{x^i},
 $$
 then the condition $[Y,X]= \lambda\, X$ implies that $\lambda=-X(Y^0)$, as a consequence of the explicit form (\ref{repvf}) of the representative vector field $X$. In this non-autonomous case, if either $M$ is an Euclidean space or choosing a local chart in $M$,  we have the associated  the volume form in $\mathbb{R}\times  M $ given by  $\Omega=dt\wedge dx^1\wedge\cdots\wedge dx^n$ and then ${\rm div\,}(X)$ and ${\rm div\,} (Y)$ take the form in such coordinates 
 $${\rm div\,}(X)=  \sum_{i=1}^n\pd{X^i}{x^i},\qquad\qquad {\rm div\,} (Y)=\pd{Y^0}{t}+  \sum_{i=1}^n\pd{Y^i}{x^i}.$$
  
 The results corresponding to Theorem \ref{divycom}, having in mind the new concept of symmetry is, according to {\it i)} of Theorem \ref{divycomR}:
 \begin{theorem}\label{nonaut} i)  If  the vector field $X$  is divergence-free and  the vector field $Y$ is an infinitesimal symmetry of the one-dimensional distribution generated by  $X$,   then  the function $\textrm{div\,}(Y)-X(Y^0) $
is a constant of the motion for $X$.

ii)  If the function  $R$ is a Jacobi multiplier  for the vector field $X$ and the vector field $Y$ is an infinitesimal symmetry  of the one-dimensional distribution 
generated by $X$,
 then the function 
\begin{equation}
I={\rm div\,}(Y)+Y(\log R)-X(Y^0) \label{fiI3}
\end{equation}
is a constant of the motion for $X$.
\end{theorem}
\hfill$\Box$

In the particular case of $M=T^*Q$, if we choose a chart for $\mathbb{R}\times T^*Q$ induced from a local chart for $Q$ and the local expression of the vector field 
$X$ is 
	$$X= \pd{}t +\sum_{i=1}^n\left(Q^i(t,q,p)\pd{}{q^i}+P_i(t,q,p)\pd{}{p_i}\right),$$ 
	which corresponds to a non-autonomous system as (\ref{asinps}) but where now the functions $Q$ and $P$ also depend on $t$,
if the coordinate expression of a vector field $Y$ on $\mathbb{R}\times T^*Q$ is 
$$Y=\sigma\, \pd{} t+ \sum_{i=1}^n\left(\xi^i\,\pd{}{q^i}+ \eta_i\, \pd{}{p_i}\right),$$ then,
as
$$[Y,X]= - (X(\sigma)\, \pd{}{t}+\sum_{i=1}^n\left((Y(Q^i)-X(\xi^i))\pd{}{q^ i}+(Y(P_i)-X(\eta_i))\pd{}{p_i}\right),
$$the conditions for $Y$ to be a normalizer of $X$, $[Y,X]=h\, X$,  are $h=-X(\sigma)$ and $Y(Q^i)- X(\xi^i)=-X(\sigma)\, Q^i$  together with  $Y(P_i)- X(\eta^i)=-X(\sigma)\, P_i$, and in thus case the associated constant of motion, for each Jacobi multiplier $R$  for $X$ with respect to the volume form $dt\wedge dq^1\wedge\cdots\wedge dq^n\wedge dp_1\wedge\cdots dp_n$  
  is 
$$I=\frac 1 R \pd{(R\sigma)}{t}+\frac 1 R \sum_{i=1}^n\left(\pd{(R\xi^i)}{q^i}+ \pd{(R\eta_i)}{p_i}\right)-X(\sigma).
$$
what shows that the results indicated in \cite{GW14} are but particular cases of the more general theory developed here.

\section{Hojman symmetry for nonautononous systems of second-order differential equations}

A  non-autonomous systems of second-order differential equations,
  \begin{equation}
  \frac{d^2x^i}{dt^2}=F^i(t, x ,\dot x ),  \quad i=1,\ldots,n,\label{secnautsyst}
   \end{equation}
   has an associated system of $(2n+1)$ first-order differential equations 
    \begin{equation}\left\{
    \begin{array}{rcl}
   {\displaystyle \frac{dt}{ds}}&=&  1\\
  {\displaystyle \frac{dx^i}{ds}}&=&v^i\\   {\displaystyle\frac{dv^i}{ds}}&=&F^i(t,  x ,v)
  \end{array}\right. \quad i=1,\ldots,n,\label{2ndautsyst}
   \end{equation}
which is a particular case of (\ref{nautsyst}) and it  is therefore geometrically described by a one-dimensional distribution $\mathcal{D}_\Gamma$ generated by a vector field $\Gamma \in \mathfrak{X}(\mathbb{R}\times T\mathbb{R}^n)$,
according to (\ref{repvf}), given by 
  \begin{equation}
  \Gamma= \pd{}{t }+\sum_{i=1}^n\left( v^i \pd{}{x^i }+ F^i(t,x,v)
\pd{}{v^i}\right).
 \label{tdnsodevf}
    \end{equation}
    
    Moreover,  in the geometric approach $\ \mathbb{R}^{n}$ is replaced by a $n$-dimensional manifold $Q$ 
and (\ref{tdnsodevf}) is the local expression of a vector field in the so-called  evolution space $E=\ \mathbb{R} \times TQ$.  Remark that $\pi_1:\ \mathbb{R} \times Q\to \ \mathbb{R} $
defines a trivial bundle and the jet-bundle of its sections is $J^1(\pi_1)= \mathbb{R} \times TQ$. 

This vector field $\Gamma$ can be characterized by the vanishing of the $2n$ 1-forms   in $ \mathbb{R} \times TQ$
\begin{equation}
\left\{ \aligned \theta^i&=dx^i- v^i \,dt,\\     
        \Psi ^i&=dv^i- F^i(t,x,v) \,dt , \endaligned  \right.\quad   i=1,\ldots 
,n,\label{distforms} \end{equation} 
together with the condition $\<dt,\Gamma>=1$.

The first $n$   1-forms $\theta^i$ are called  contact 1-forms. Any other representative of the distribution $\mathcal{D}_\Gamma$  will be of the form 
   \begin{equation}
  \bar \Gamma=\lambda(t,x,v)   \Gamma=\lambda(t,x,v)\left(
 \pd{}{t }+\sum_{i=1}^n\left( v^i \pd{}{x^i }+ F^i(t,x,v)\pd{}{v^i}\right)\right),
\label{tdsodevf}
    \end{equation}
    with $\lambda(t,x,v)$ a nonvanishing function. Vector fields of type (\ref{tdsodevf}) are called  {{\small{SODE}} vector fields while those of type  (\ref{tdnsodevf}) are said to be Newtonian  {{\small{SODE}}   (hereafter shortened as {\small{NSODE}}) vector fields. 

We follow notation of \cite{MC77} and we therefore refer to such a paper for definitions.  More details can be found in \cite{CFM91,CFM92}. 
An important point is that the set of  1-forms
 \begin{equation}
\{dt,    \theta^i=dx^i-v^i\,dt,
\Psi^i=dv^i-F^i(t,x,v)\,dt\}  ,       \label{adaptedbasisforms}
\end{equation}
 is a local basis of 1-forms in $ \mathbb{R}\times TQ$   which  is the  dual basis of  vector fields in  $\mathfrak{X}( \mathbb{R}\times TQ)$,
given by  
 \begin{equation}
\left\{ \Gamma, \pd{}{x^i},\pd{}{v^i} \right\} .    \label{adaptedbasisvf} \end{equation} 
Note that while the contact forms only depend on the evolution space, the $n$ 1-forms $\Psi^i$ depend also on the selected NSODE vector field $\Gamma$. 
Particular examples for this case of second order differential equations    were studied in \cite{SH92a, FGGJPA27,FGGLNC19}. 

We only want to remark here that the Lagrangian formalism is defined for a function $L\in C^\infty(\mathbb{R} \times TQ)$ by means of a $(1,1)$-tensor in the evolution space $E$, called vertical endomorphism, whose local coordinate expression is  
 \begin{equation}
 S=\sum_{i=1}^n\pd{}{v^i}\otimes         \theta^i=\sum_{i=1}^n\pd{}{v^i}\otimes 
(dx^i-v^ i\,dt).\label{defStdep}
\end{equation}  
This may be used to define the 1-form $\Theta _L= L\, dt +dL\circ   S$ in $E$, 
 $$\Theta _L= L\, dt +\<dL\circ   S,\G>\, dt+\sum_{i=1}^n
\left\langle dL\circ  
 S,\pd{}{x^i}\right\rangle\theta^i + \sum_{i=1}^n\left\langle dL\circ  
S,\pd{}{v^i}\right\rangle\Psi^i, $$
and as $S(\G)=0$  and $S(\partial/\partial v^i)=0$, $\forall i=1,\ldots,n$, we find the coordinate expression of $\Theta _L$,
\begin{equation}\Theta_L= L\,  dt+\sum_{i=1}^n \pd L{v^i} \,\theta^i.     \label{Cartan2}
\end{equation}

The exterior differential $d\Theta_L$ is  used to define the 2-form  $\Omega_L=-d\Theta_L$, called Cartan 2-form. When $\Omega_L$ is of rank $2n$ the Lagrange function
 is called regular. In this case, there is only one vector field $\Gamma$ such that $i(\Gamma)\Omega_L=0$ and $\<dt,\Gamma>=1$,  which is the dynamical vector
  field defined by the regular function $L$.
  
 The coordinate expression of the Cartan 2-form is:
 \begin{equation}\Omega_L=-d\Theta_L=\sum_{i,j=1}^n\left(-A_{ij} \theta^i \wedge \theta^j+W_{ij} \theta^i \wedge \Psi^j\right) , \label{exprCartan2form}\end{equation}
 with the functions $A_{ij}$ and $W_{ij}$ being given by
 \begin{equation}\qquad A_{ij}=\pd{^2L}{x^i\partial v^j},\quad  W_{ij}=\pd{^2L}{v^i\partial v^j},\quad i,j=1,\ldots,n. \label{defAW}
 \end{equation}
 We recall that the dynamical vector field $\Gamma$ satisfies the two conditions  $i(\Gamma) \theta^i=0$ and $ i(\Gamma)\Psi^j=0$ and, because of this, it satisfies also  the equation $i(\Gamma) \Omega_L=0$.
Furthermore, 
 $(\W _L )^{\wedge  n}$ is a real multiple of $(\det W_{ij})\,  \theta^1\wedge \ldots   \wedge  \theta^n 
\wedge \Psi^1\wedge \ldots \wedge  \Psi^n$.  Note also that as $dt\wedge \theta^i=dt\wedge dx^i$ and  $dt\wedge \Psi^i=dt\wedge dv^i$, we have that $dt\wedge (\Omega_L)^{\wedge n}=(\det W_{ij})\bar\Omega$  where $\bar\Omega$ denotes  the volume form in the evolution space $ \bar\Omega=dt\wedge dx^1\wedge\cdots\wedge dx^{n}\wedge dv^1 \wedge\cdots\wedge dv^{n}$. But  as $\mathcal{L}_\Gamma  \Omega_L=d(i(\Gamma) \Omega_L)=0$ and $\mathcal{L}_\Gamma dt =0$, we have that the vector field $\Gamma$ is 
divergence-free with respect to the volume form  $ dt\wedge(\Omega_L)^{\wedge n}$,     
Consequently  the determinant of the matrix with elements $W_{ij}$ is a Jacobi multiplier with respect to the volume form in the evolution space 
$\bar \Omega=dt\wedge dx^1\wedge\cdots\wedge dx^{n}\wedge dv^1 \wedge\cdots\wedge dv^{n}$.

Therefore, this allows us to find as a particular case the result of \cite{Lutzky95}, because if the vector field 
with coordinate expression  
$$Y=Y^0(t,x,\dot x)\pd{}t+ \sum_{i=1}^n\left(Y^i(t,x,\dot x)\pd{}{x^i}+\bar Y^i(t,x,\dot x)\pd{}{v^i}\right),
 $$
 is a symmetry of the vector field $\Gamma$ derived from a Lagrangian $L$, then, according to (\ref{fiI})  the function 
 $$
I={ \rm div\,}Y+ Y(\log W)
 $$
is such that $ \Gamma(I)=0$, where in this case the divergence should be computed with respect to the volume form $\Omega$: 
$${ \rm div\,}Y=\pd{Y^0}t+\sum_{i=1}^n\left(\pd{Y^i}{x^i}+\pd{\bar Y^i}{v^i}\right).
$$

Having in mind that 
$$[Y,\Gamma]=- \Gamma(Y^0)\pd{}t+\sum_{i=1}^n\left( (\bar Y^i-\Gamma(Y^i))\pd{}{x^i}+(Y(F^i)-\Gamma(\bar Y^i))\pd{}{v^i}\right),
$$
we see that $[Y,\Gamma]=0$ if and only if $ \Gamma(Y^0)=0$, 
          $\bar  Y^i=\Gamma(Y^i)$ and $\Gamma(\Gamma(Y^i))=Y(F^i)$. Furthermore,  $Y$ is an infinitesimal symmetry of the distribution generated by $\Gamma$,  i.e. there exists a function $h\in C^\infty(M)$ such that $[Y,\Gamma]=h\, \Gamma $,  if and only if $h=-\Gamma(Y^0)$, $\bar  Y^i=\Gamma(Y^i)+h\, v^i$ and $Y(F^i)=\Gamma(\bar Y^i)+h\, F^i$. Therefore  we obtain 
 $$\bar  Y^i=\Gamma(Y^i)- \Gamma(Y^0)\, v^i ,
 $$ 
 and 
 $$Y(F^i)=\Gamma(\Gamma(Y^i))-2\Gamma(Y^0)\, F^i- \Gamma(\Gamma(Y^0))v^i .
 $$
The two particular cases  $Y^0\equiv 0$ and  
$Y^i\equiv 0$ were studied in  \cite{ZCh}.

Particularly interesting cases are those for which the functions $Y^0$ and $Y^i$ do not depend on velocities because in this case $\Gamma (Y^0)$ and $\Gamma (Y^i)$ do not depend on the choice of the NSODE vector field $\Gamma$ but only of its NSODE character. On the other side, these vector fields play a relevant role in the geometric formulation of 
Noether theorem, where the symmetry of the dynamical vector field is analysed in terms of symmetry under this kind of vector fields of the Lagrangian itself.  We first recall some basic notions of geometry  of the evolution space.

In the study of nonautonomous systems of second-order differential equations the contact 1-forms $\theta^i$ play a relevant role, because they can be used to characterize curves 
$\gamma $ in $\mathbb{R}\times TQ\to \mathbb{R}$ that are prolongations of sections of the trivial bundle $\mathbb{R}\times Q$ over $\mathbb{R}$: In fact, such curves are characterized by $\gamma^*\theta^i=0$, for $i=1,\ldots,n$. 

Accordingly, given a vector field $X\in \mathfrak{X}(\mathbb{R}\times Q)$ there exists one vector field   $X^{(1)}\in  \mathfrak{X} (\Bbb R\times TQ)$
such that it is projectable on $X$ and preserves the contact distribution. If 
the coordinate expression of $X\in \mathfrak{X}(\mathbb{R}\times Q)$ is
 \begin{equation}
X=X^0 (t,x)\pd{}{t }+\sum_{i=1}^n\left( X^i(t,x)\pd{}{x^i}\right),\label{XinRxQ}\end{equation} 
then a vector field 
$X^{(1)}\in \X (\Bbb R\times TQ)$   projectable on 
$X$, must be of the form  
\begin{equation}
X^{(1)}=X^0 (t,x)\pd{}{t}+\sum_{i=1}^n\left( X^i(t,x)\pd{}{x^i}+
\bar X^i (t,x,v)\pd{}{v^i}\right), 
        \label{XinRxQ}
\end{equation}
where, choosing an arbitrary NSODE vector field and the corresponding associated bases of 1-forms and vector fields,  as 
 $$\mathcal{L}_{X^{(1)}}\theta^i = \sum_{j=1}^n
\left(   \pd{X ^i}{x^j}- v^i \pd{X^0  }{x^j}\right)\,
\theta^j+( \Gamma X ^i-\bar X^i -v^i \, \Gamma X^0 )\, dt,$$
this  shows that in order to the set of   contact 1-forms  be preserved by $X^{(1)}$, it must be of the form 
  \begin{equation}
\bar X^i =\Gamma(X ^i)-v^i\, \Gamma(X^0),   \quad i=1,\ldots, n.\label{preservcond}
       \end{equation}
Note that the last relations are valid for any NSODE $\Gamma$.  

Next we consider a particular case that contains some of the results obtained in \cite{ZCh}.     

In the particular case of a vector field $X^{(1)}$ that is a symmetry of the one-dimensional distribution generated by a NSODE  vector field $\Gamma$ determined by a Lagrangian $L$,
as $\det W$ is a Jacobi multiplier for $\Gamma$ the Hojman constant of the motion according to (\ref{fiI2})  is  given by 
  \begin{equation}
I= 2\sum_{i=1}^n\left( \pd {X^i}{x^i}- v^i\pd{X^0}{x^i}\right)-n\, \Gamma(X^0)+X^{(1)}(\log (\det W)), \label{fiI4}
\end{equation}
because 
$${ \rm div\,}(X^{(1)})= \pd{X^0} t+ \sum_{i=1}^n\left(\pd{X^i}{x^i}+\pd{\bar X^i}{v^i}\right),
$$
and as $X^0$ and $X^i$ do not depend on the velocities, 
$$
\pd{ (\Gamma(X^0))}{v^i}=\pd{X^0}{x^i},\qquad \pd{ (\Gamma(X^i))}{v^i}= \pd{X^i}{x^i},\,
$$ 
and then,       
$${ \rm div\,}(X^{(1)})= \pd{X^0} t+ \sum_{i=1}^n\left(\pd{X^i}{x^i}+\pd{(\Gamma(X^i)-v^i\, \Gamma(X^0))}{v^i}\right),$$
and therefore
$${ \rm div\,}(X^{(1)}) = \pd{X^0} t+\sum_{i=1}^n\left(2\pd{X^i}{x^i}-v^i\pd{X^0}{x^i}-\Gamma(X^0)\right), 
$$        
 which can be rewritten as 
 $${ \rm div\,}(X^{(1)}) = 2\sum_{i=1}^n\left(\pd{X^i}{x^i}-v^i\pd{X^0}{x^i}\right)-(n-1)\Gamma(X^0), 
$$
and, as $\det W$ is a Jacobi multiplier,  the constant of the motion corresponding to (\ref{fiI3})    is  (\ref{fiI4}), 
in agreement with the  above mentioned results in \cite{ZCh}. 

\section{Conclusions and outlook}

It has been shown how the use of basic results of the differential geometry concerning the concept of divergence of a vector field, as well as some properties relating Lie 
derivatives of forms with respect to couples of vector fields, allows us to establish in Theorem \ref{divycom} a method to find a first-integral for a divergence-free vector field
 when a symmetry vector field is known. The theory is generalised to include  a more general vector field when a Jacobi multiplier for such given vector field is known in 
 Theorem \ref{hojman}. Theorem  \ref{divycomR} presents the extension of the results 
of these Theorems when a normalizer vector field instead of a symmetry vector is known. This generalization is very relevant for the case of a nonautonomous system. The main interest of these Theorems is that they present in an intrinsic way a general
 result that contains as particular cases   many expressions of first-integrals or constants of motion derived by ad-hoc methods in many branches of physics. This makes very interesting 
 this unifying way of finding them using geometric tools of modern differential geometry. Several examples of applications to general Hamiltonian systems,  second order differential equations and  Lagrangian systems, both in the autonomous and the nonautonomous cases, have also been exhibited.

\end{document}